%% file: root.tex
\documentclass[letterpaper, 10 pt, conference]{ieeeconf}  

\IEEEoverridecommandlockouts

\usepackage{algorithm}
\usepackage{algorithmic}
\usepackage{booktabs}
\usepackage{array} 
\usepackage{makecell}
\usepackage{subcaption}
\usepackage{amsfonts}
\usepackage{cite}
\usepackage{graphics} 
\usepackage{graphicx}
\usepackage{epsfig} 
\usepackage{mathptmx} 
\usepackage{times} 
\usepackage{amsmath} 
\usepackage{amssymb}  
\usepackage{tabularx}
 \usepackage{multirow}
\usepackage{hyperref}
\usepackage{authblk}
\usepackage[dvipsnames]{xcolor}

\title{\LARGE \bf Design and Implementation of a Scalable Clinical Data Warehouse for Resource-Constrained Healthcare Systems}

\begin{document}

\author{Shovito Barua Soumma$^{1,3}$, Fahim Shahriar$^{2,3}$, Umme Niraj Mahi$^4$, Md Hasin Abrar$^3$, \\Md Abdur Rahman Fahad$^{3,5}$,  ASM Latiful Hoque$^3$ \\
\small
$^1$College of Health Solutions, Arizona State University, USA\\
$^2$Computer Science Department, University of Minnesota Duluth, USA\\
$^3$Computer Science and Engineering Department, Bangladesh University of Engineering and Technology, Bangladesh\\
$^4$Computer Science and Engineering Department, Khulna University of Engineering and Technology, Bangladesh\\
$^4$Computer Science Department, Missouri State University, USA\\
\thanks{
\vspace{-5mm}
$^*$Corresponding author. Email: 
{\color{blue}shovito@asu.edu}}
}

\maketitle
\thispagestyle{empty}
\pagestyle{empty}

\begin{abstract}
Centralized electronic health record (EHR) repositories are critical for advancing disease surveillance, public health research, and evidence-based policymaking. However, developing countries face persistent challenges in achieving this due to fragmented healthcare data sources, inconsistent record-keeping practices, and the absence of standardized patient identifiers. These issues hinder reliable record linkage, compromise data interoperability, and limit scalability—obstacles exacerbated by infrastructural constraints and privacy concerns. To address these barriers, this study proposes a scalable, privacy-preserving clinical data warehouse, NCDW, designed for heterogeneous EHR integration in resource-limited settings deployed on institutional servers and tested with 1.16 million clinical records. The framework incorporates a wrapper-based data acquisition layer for secure, automated ingestion of multisource health data and introduces a space-efficient Soundex algorithm to resolve patient identity mismatches in the absence of unique IDs. A modular data mart is designed for disease-specific analytics, demonstrated through a dengue fever case study in Bangladesh, integrating clinical, demographic, and environmental data for outbreak prediction and resource planning. Quantitative assessment of the data mart underscores its utility in strengthening national decision-support systems, highlighting the model’s adaptability for infectious disease management. Comparative evaluation of database technologies reveals NoSQL outperforms relational SQL by 40-69\% in complex query processing, while system load estimates validate the architecture’s capacity to manage 19 million daily records (34TB over 5 years). The framework can be adapted to various healthcare settings across developing nations by modifying the ingestion layer to accommodate standards like ICD-11 and HL7 FHIR, facilitating interoperability for managing diverse infectious diseases (i.e., COVID, tuberculosis, malaria). 
Code is available: \href{https://github.com/shovito66/ncdw}{\color{blue}https://github.com/shovito66/ncdw}.

\indent \textit{Index Terms}— Dengue, Datamart, Health Informatics, Data Mining, Warehouse, Electronic Health Records
\end{abstract}

\input{tex/intro_related}

\input{tex/proposed_architechture}

\input{tex/results}

\input{tex_generalized/limitations}
\input{tex_generalized/conclusion}

\vspace{-2mm}
\bibliographystyle{IEEEtran}
\bibliography{sample.bib}

\end{document}

%% file: tex/intro_related.tex
\section{Introduction \& Related Works}




The accelerated digitization of healthcare services globally has led to an unprecedented surge in Electronic Health Records (EHRs) and other clinical data. In principle, data warehouses can integrate diverse healthcare information into a centralized repository, which forms the basis for advanced analytics, population health management, and evidence-based policymaking \cite{evans2016electronic}. Developed nations have successfully implemented Clinical Data Warehouses (CDWs) to enhance healthcare delivery and research. For instance, a study examining 32 French regional and university hospitals found that 14 had fully operational CDWs, with implementations accelerating since 2011 \cite{France}. In the United States, institutions have developed CDWs to support clinical decision-making and secondary data use by integrating heterogeneous data sources \cite{Usa}. However, in developing countries such as Bangladesh, where healthcare systems are fragmented, record-keeping is inconsistent, and infrastructure is limited, fully reaching this potential remains a challenge~\cite{wisniewski2003development, 7400708}.

A National Clinical Data Warehouse (NCDW) offers a viable approach to consolidating multi-institutional data into a privacy-preserving environment. By unifying patient information, laboratory results, and clinical observations, an NCDW can enhance large-scale health analytics and support more informed policy decisions \cite{adewole2024systematic}. Data warehouse architectures often incorporate specialized data marts tailored to specific domains or diseases by integrating clinical, geographic, and environmental data~\cite{heinonen2019data}. For instance, in dengue-endemic regions like Bangladesh, where seasonal outbreaks pose significant healthcare challenges, a dedicated dengue-focused data mart can enhance the opportunity to detect outbreaks, plan resource allocation, and conduct epidemiological research \cite{kayesh2023increasing,mutsuddy2019dengue}. Data mining techniques such as clustering and association can also uncover hidden patterns in health data, aiding outbreak detection~\cite{han2012data,khamesian2025type1diabetesmanagement, 10054906}.

The choice of SQL (e.g., PostgreSQL) vs. NoSQL (e.g., HBase) is also critical for managing healthcare data. SQL ensures ACID (atomicity, consistency, isolation, durability) compliance, while NoSQL offers better scalability for semi-structured data. A recent study in Bangladesh suggests hybrid approaches improve performance for large-scale health analytics~\cite{khan2015development}.

However, several key challenges limit the effectiveness of NCDWs in resource-constrained settings. Initially, the absence of unique patient identifiers, often due to low literacy rates, variable name spellings, and minimal official documentation, complicates record linkage. Secondly, selecting an appropriate database platform (e.g., SQL vs. NoSQL) proves critical for efficiently storing and querying large-scale health data under infrastructural constraints. Thirdly, integrating environmental or demographic data remains essential for analyzing and predicting infectious diseases such as dengue, yet many existing solutions lack a mechanism to incorporate these external factors. Lastly, machine learning (ML) algorithms for real-time outbreak detection require consistent, high-quality data, which can be difficult to obtain in developing contexts.

In earlier work \cite{ncdwProblemsIssues}, existing national health data management systems have been analyzed, along with the design and limitations of Bangladesh’s NCDW, and the need to strengthen its performance, security, and integration of heterogeneous data has been explored. Our previous works~\cite{mia2022privacy,ksrl} proposed a privacy-preserving NCDW architecture that leveraged the key-based algorithm to anonymize patient data while enabling secure linkages across fragmented sources. Although this solution addressed fundamental privacy concerns, limitations were identified in areas such as scalability, interoperability, and the incorporation of external datasets for comprehensive analysis. To overcome these challenges, this study presents an enhanced NCDW architecture with the following key improvements:
\begin{enumerate}
    \item \textbf{A wrapper-based secure data integration from heterogeneous sources.}
    \item \textbf{ External data (i.e., environmental and demographic)  integration in our CDW.}
    \item \textbf{A name-matching algorithm for improved record linkage.}
    \item \textbf{A specialized data mart for disease-specific analysis.}
    \item \textbf{SQL vs. NoSQL performance evaluation to enhance system scalability and performance.}
\end{enumerate}


%% file: tex/proposed_architechture.tex
\section{Proposed Architecture}
\label{sec:methodology}

\begin{figure*}[ht]
    \centering
    \includegraphics[width=0.9\linewidth,trim={2 3 5 5},clip]{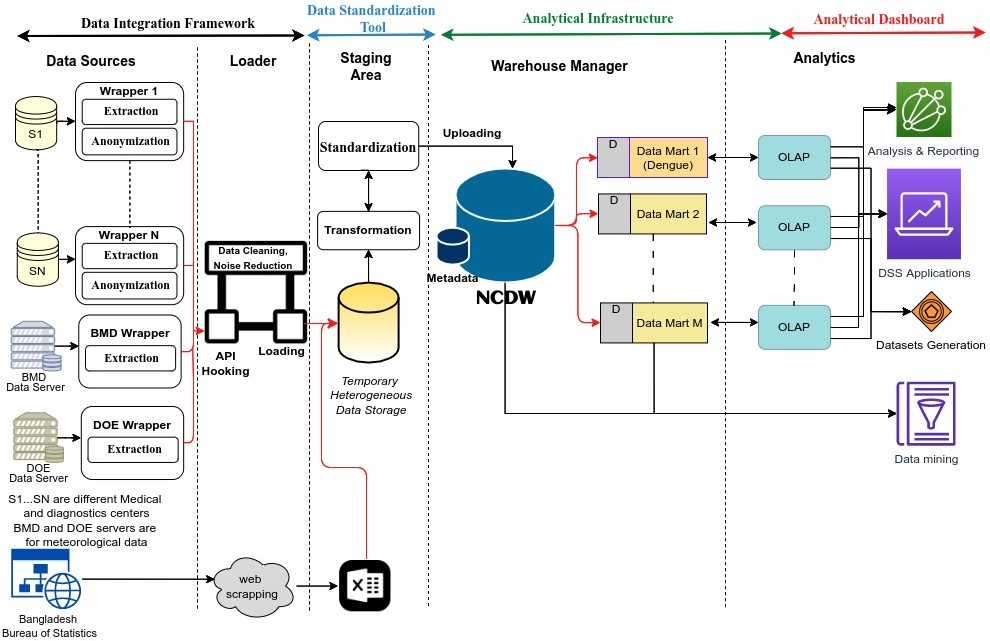}
    \caption{Wrapper-based architecture of NCDW supporting disease-specific data mart.}
    \label{fig:ncdw_architecture}
    \vspace{-3mm}
\end{figure*}

The proposed National Clinical Data Warehouse (NCDW) architecture, shown in Fig.~\ref{fig:ncdw_architecture}, can be divided into four stages: (1) Data Integration, (2) Data Standardization, (3) Centralized Data Management, and (4) Analytical Infrastructure and Dashboard. 

\subsection{Data Integration}
\label{subsec:data_integration}

The data integration process employs a \textbf{wrapper-based framework} to collect data from healthcare providers, environmental agencies, and population statistics. The data is anonymized using the Key-Based Secured Record Linkage (KSRL) algorithm~\cite{ksrl}, which replaces patient identifiers with pseudonymous keys, and then cleaned to resolve errors and reduce noise. Finally, the processed data is loaded into the CDW's temporal storage using an \textbf{API-based plugin} for standardization and transformation.

\subsection{Data Standardization}
\label{subsec:data_standardization}


The NCDW follows global data standards for accuracy, consistency  and  interoperability in healthcare records. The HL7 Clinical Research Functional Profile (CRFP) for EHR facilitates structured data exchange, while SNOMED CT and LOINC define a common language for clinical terms and laboratory tests. ICD-11 classifies diseases, UCUM standardizes units, and NHS Pathology categorizes lab tests~\cite{NCVHS2018,ICD-11}. Transformation rules standardize timestamps and environmental data (e.g., temperature, humidity, rainfall)~\cite{batini2009methodologies}, and the Intelligence Data Interfacing Model (IDIM) prepares data for warehouse loading. Batch processing merges and loads the data into the multidimensional models of the central warehouse for parallel processing.

\subsection{Centralized Data Warehousing and Management}
\begin{figure}[!ht]
    \centering
    \includegraphics[width=1.0\linewidth,trim={15 15 15 15},clip]{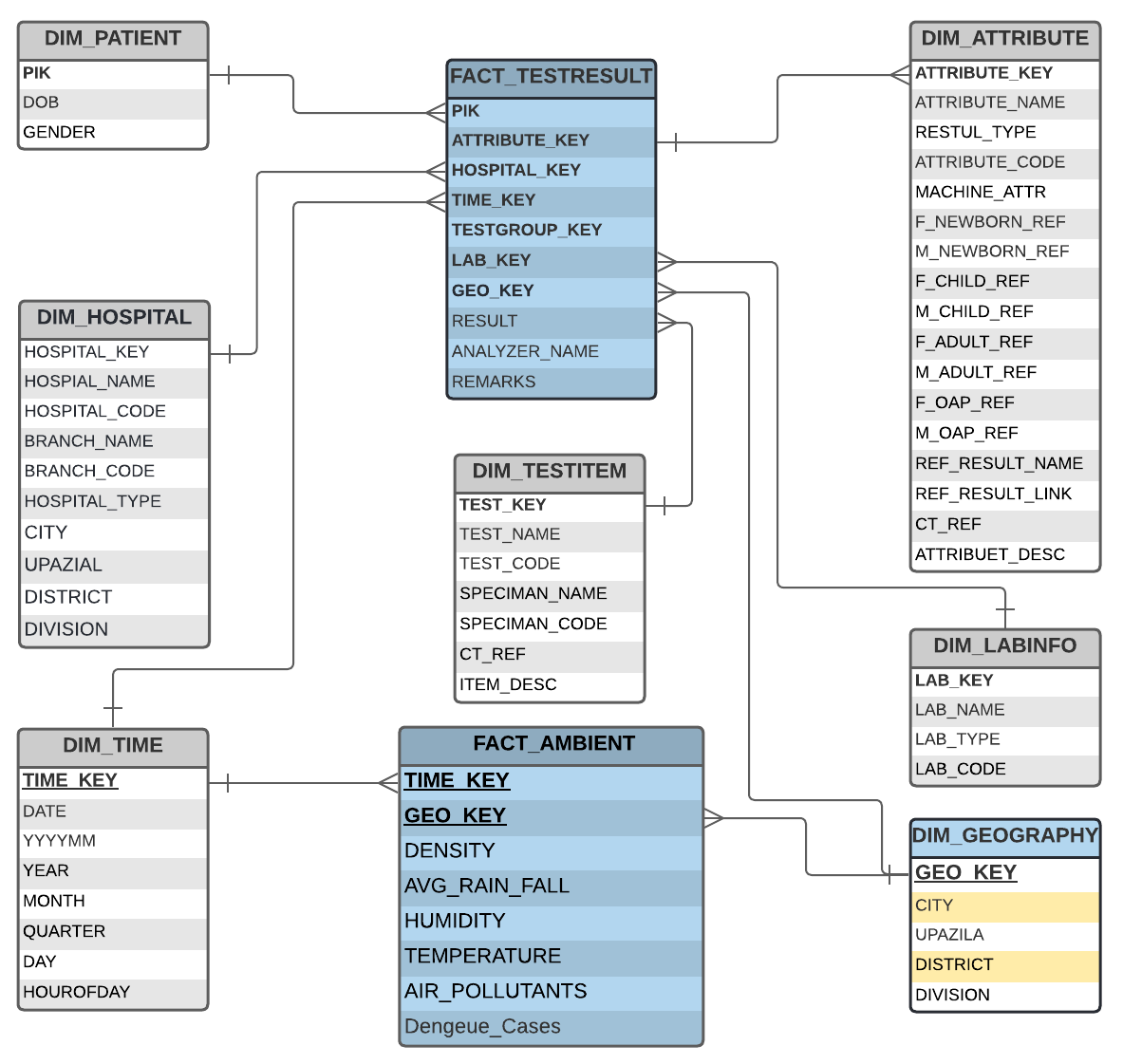}
    \caption{Star schema of NCDW supporting dengue data mart.}
    \label{fig:star_schema}
    \vspace{-2.5mm}
\end{figure}

\vspace{-1.5mm}
The NCDW employs a star schema architecture, as shown in Fig.~\ref{fig:star_schema}, a widely used analytical data modeling approach for its computational efficiency~\cite{Nugawela, Insights_clinical, 4743972}. The schema includes two fact tables: TESTRESULT and AMBIENT, along with eight dimension tables. The newly added dimension, GEOGRAPHY, enhances dengue outbreak detection by incorporating the attributes CITY, UPAZILA, DISTRICT, and DIVISION linked by a unique GEOKEY generated during transformation. The AMBIENT fact table connects to the TIME and GEOGRAPHY dimensions, recording five key measures: density, average rainfall, humidity, air pollutants, and percentage of positive dengue patients. The dynamic schema leaves scope for future extensions, such as adding genomic or molecular data.

Dimension identifiers, like healthcare key, lab key, and attribute key, are generated as five-digit IDs to minimize fact table size and computational complexity. The TIME key is represented as UNIX format (e.g., 847493700 for 8th Nov 1996, 3:55 PM), while the Patient Identifier Key (PIK) is anonymized for privacy. Table~\ref{tab:new_tables} summarizes the newly added tables.

\begin{table}[ht]
    \centering
    \caption{Details of newly added fact and dimension tables.}
    \label{tab:new_tables}
    \resizebox{\linewidth}{!}{%
    \begin{tabular}{ccp{3cm}}
        \toprule
        \textbf{Type} & \textbf{Table Name} & \textbf{Description} \\ 
        \midrule
        Fact          & AMBIENT             & Detection and prediction of dengue outbreak \\ 
        Dimension     & GEOGRAPHY           & Logs ambient data and geographical attributes \\ \bottomrule
    \end{tabular}%
    }
\end{table}

\begin{figure}[!ht]
    \centering
    \includegraphics[width=0.77\linewidth]{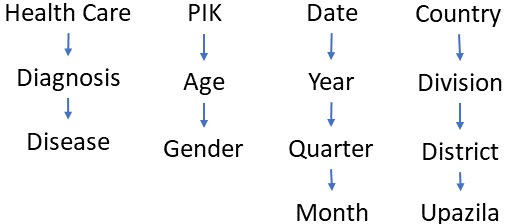}
    \caption{CUBE hierarchy in disease analysis.}
    \label{fig:cube}
\end{figure}

\begin{figure}[!ht]
    \centering
    \includegraphics[width=1.0\linewidth]{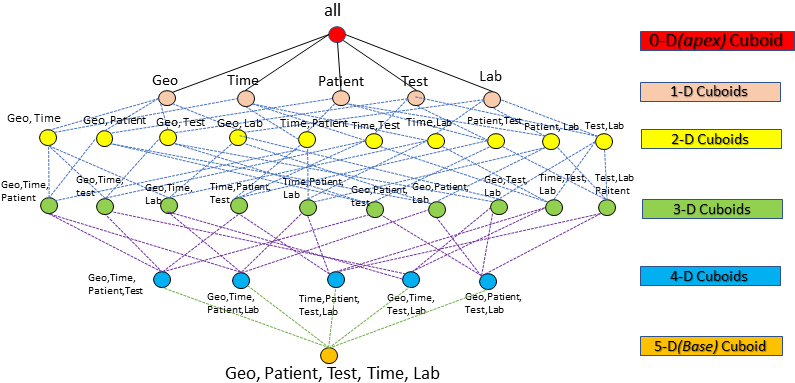}
    \caption{Clinical test data cube.}
    \label{fig:testdatacube}
    \vspace{-6.5mm}
\end{figure}

\vspace{-2mm}
\textit{Cube Hierarchy and Aggregation:} As illustrated in Fig.~\ref{fig:cube}, aggregated data from the schema enables the creation of multi-dimensional cubes for diagnosis and disease analysis. The 5D clinical test data cube shown in Fig.~\ref{fig:testdatacube} supports NCDW development, with a cube hierarchy ranging from the 0-D apex cube, summarizing national health statistics, to the 4-D base cube for granular data. The system precomputes common analytics, such as patient counts by diagnosis, procedure tracking, and retesting statistics, ensuring fast and efficient query responses. These preaggregated datasets enhance knowledge discovery, prediction, and dataset generation, aligning with the warehouse's goal of enabling real-time insights.

To identify the best storage solution for NCDW, PostgreSQL and HBase were compared using 3D and 4D data cubes (100K–1M records) based on execution time and efficiency. PostgreSQL suits structured clinical data needing ACID compliance, while HBase offers scalability and fast writes for environmental and genomic data.

\subsection{Analytical Infrastructure and Dashboard}
\label{subsec:analytical_infrastructure}
The analytical infrastructure integrates Online Analytical Processing (OLAP) and data mining to detect, predict, and analyze disease outbreaks. Using dengue as a proof-of-concept, the system demonstrates its ability to analyze climate-disease correlations and forecast outbreaks. Data is processed and standardized using the Intelligence Data Interfacing Model (IDIM) and batch-loaded into the NCDW, enabling parallel processing and automated generation of disease-specific data marts for infectious diseases such as COVID-19, Malaria, Dengue, and other infectious and parasitic diseases.

The dashboard provides a read-only interface with interactive visualizations, real-time reports, and predictive analytics. It forecasts disease spread and severity, examines seasonal and geographical climate-disease relationships, and evaluates medical capacity and intervention readiness. These results support evidence-based planning and public health interventions, with the system designed for scalability across other diseases. 



\subsection{Dataset and Experimental Setup}
\label{subsec:dataset_description}

The NCDW integrates four datasets: \textbf{Clinical}, \textbf{Dengue}, \textbf{Environmental}, and \textbf{UrbanStats}. Clinical dataset consists of anonymized health records of 9443 patients, while the Dengue dataset includes 70,049 test records and 101,658 records for 69633 anonymous patients, covering all districts. Environmental and demographic data from Bangladesh Meteorological Department (BMD),
Department of Environment (DOE),
and Bangladesh Bureau of Statistics (BBS)
provide weather parameters and population density for disease analysis. Table~\ref{tab:datasets_summary} provides a summary of these datasets.

\begin{table}[]
    \centering
    \caption{Summary of datasets used in NCDW.}
    \label{tab:datasets_summary}
    \begin{tabular}{p{1.5cm}p{1.5cm}p{4cm}}
        \toprule
         \textbf{Dataset (\#records)}         & \textbf{Sources}              & \textbf{Key Attributes}  \\
         \hline
         Clinical (1161654)       & 13 Hospitals           & Anonymized patient records, test results, age, gender \\
         \hline
         Dengue\hspace{0.5cm}  (70049)          & 6 Hospitals (2016–2022)       & Test outcomes, gender(39801 male, 30246 female, 2 other),+Ve cases(16510), geographical coverage across all districts \\ 
         \hline
         Environmental &  BMD, DOE & rainfall, humidity, air quality, temperature\\ 
         \hline
         UrbanStats        & BBS &Population/Building Density \\
         \bottomrule
    \end{tabular}
\end{table}

To validate the proposed architecture, a prototype was deployed on a server at IICT, BUET, with a 2×4-core CPU, 150 GB HDD, and 32 GB RAM. The web-based interface enabled interactive analysis via maps and charts, while PostgreSQL managed the backend. A subject-oriented data mart, \emph{DENGUE}, was derived for specialized dengue analysis.









\subsection{Name Matching for Data Consistency}

In developing countries like Bangladesh, patient identification is challenging due to inconsistent healthcare data, low literacy rates, and the absence of standardized identity documents. Variations in name spellings, repeated ages due to missing birth records, and high patient volumes further contribute to data inconsistencies. Limited staff expertise and rushed data entry introduce errors, making traditional record linkage methods ineffective in such environments.

\begin{algorithm}
\caption{Soundex Name Matching Algorithm. $\alpha$ is a function to retain the first character. $f$ is the mapping fuction}
\textbf{Input:} A string $S=\{S_1S_2...S_n\}$ representing a name.\\
\textbf{Output:} A four-character soundex code $C$ for $S$.

\begin{algorithmic}[1]
\label{alg:soundex}
\STATE {\bfseries Begin}
\STATE Initialize $C \gets \text{empty string}$.
\STATE Retain the first letter of $S$: $C_1 \gets \alpha(S_1)$.
\STATE Initialize $P \gets \text{MappingFunction} f(C_1)$.

\FOR{$i=2$ to $n$ in $S$}
    \IF{$f(S_i) = \emptyset$}
    \STATE \textbf{continue}
  \ELSIF{$f(S_i) \neq P$}
    \STATE Append $f(S_i)$ to $C$
    \STATE Update $P \gets f(S_i)$
  \ENDIF
\ENDFOR

\IF{$|C| < 4$}
    \STATE Append "0" to $C$ until $|C| = 4$.
\ELSIF{$|C| < 4$}
    \STATE Truncate $C$ to the first four characters.
\ENDIF

\STATE {\bfseries return} $C$
\STATE {\bfseries End}
\end{algorithmic}
\end{algorithm}
\vspace{-3mm}


\begin{table}[!h]
    \centering
    \caption{Mapping Function}
    \label{tab:map_func}
    \begin{tabular}{c|c}
    \toprule
         Value of $x$&  $\text{MappingFunction }f(x)$\\
         \midrule
         $x \in \{B, F, P, V\}$ & 1\\
         $x \in \{C, G, J, K, Q, S, X, Z\}$ &2\\
         $x \in \{D, T\}$ &3\\
         $x=L$ &4\\
         $x \in \{M, N\}$ &5\\
         $x \in \{M, N\}x = R$ &6\\
         $x \in \{A, E, I, O, U, H, W, Y\}$ &$\emptyset$ \\ 
         \bottomrule
    \end{tabular}
\end{table}
To address patient identification challenges, we proposed a name-matching algorithm, Soundex [\ref{alg:soundex}] that generates phonetic identifiers to handle spelling variations and pronunciation differences. It converts names into a fixed-length four-character code, reducing ambiguity and ensuring standardized comparison. For instance, “Sobuj Chowdhury” and “Sabuj Chaudhury” produce identical Soundex codes (S132 C360), enabling accurate record linkage. This approach minimizes data inconsistencies and improves patient record management in resource-constrained healthcare systems, as shown in Table~\ref{tab:soundex}.
\begin{table}[!h]
    \centering
    \caption{Usage of Soundex encoding process for name standardization. \textbf{*} before adjustment.}
    \label{tab:soundex}
    \setlength{\tabcolsep}{2pt}
    \begin{tabular}{c|cccc}
    \toprule
         \textbf{Inputted Name }&\textbf{1st Lette}r &\textbf{Mapped Values} & \textbf{Code*}  &\textbf{Final Code}  \\
         \midrule
         Sabuj/Sobuj/Sabuz &S &1,3,2 &S132 &S132\\
         Chowdhury/Choudhury & C &3,6 &C36 &C360\\
         Smyth/Smith/Smeth &S &5,3 &S53 &S530\\
         \bottomrule
    \end{tabular}
\end{table}

%% file: tex/results.tex
\vspace{-4mm}
\section{Result \& Discussion}
The implementation of the NCDW  successfully integrated clinical, demographic, and environmental datasets, demonstrating its ability to manage multidimensional data and support real-time healthcare analytics. Our findings can be generalized into two main types. First, we will discuss our findings on the load estimation, storage requirements, and performance comparison of our proposed system. Then we will share our dengue data mart specific findings. Together, they highlight the capacity of the system and its role in data-driven public health interventions.

\subsection{Infrastructure and Performance Findings}
This one focuses on the system’s infrastructure and performance, assessing its ability to integrate clinical datasets, manage storage, and process data efficiently.

\subsubsection{Load Estimation and Storage Requirements}
\label{subsec:load_estimation}

Estimating the daily data load and total storage requirements of the NCDW is crucial for efficient management. The daily load is calculated by aggregating record entries from various healthcare facilities, using the formula:
\begin{equation}
     R =\bar{r}\times  \left \{ \sum_{j=1}^{n} \left ( \frac{s_j*hospital\,count}{\sum_{}^{total\,bed}s} \right ) \right \}\times d(days)
\end{equation}

where $R$ represents daily record entries, $\bar{r}$ is the average records per hospital, $s_j$ denotes hospital weight based on bed count, $n$ is the total hospitals, and $d$ is the number of days.

The total NCDW size is estimated as:
\vspace{-2mm}
\begin{equation}
\vspace{-2mm}
\label{eqn2}
\text{Size} = R \times d \times S_r 
\end{equation}

where $S_r$ is the average record size in kilobytes. These calculations help in planning storage needs for different periods (e.g., one to five years), ensuring scalability through cloud-based solutions.

By applying these formulas, future data growth can be anticipated, enabling proactive infrastructure planning.

\begin{table}[!h]
    \centering
    \caption{Average number of daily records entered in a hospital over a week.}
    \label{tab:daily-avg-record}
    \setlength{\tabcolsep}{2pt}
    \begin{tabular}{c|cccccccc}
        \toprule
         \textbf{Day} &\textbf{Sun}& \textbf{Mon}& \textbf{Tue} &\textbf{Wed} &\textbf{Thu} &\textbf{Fri} &\textbf{Sat}  \\
         \textbf{Avg Record, \(r_i\)} &10072 &9976 &10132 &9931 &8973 &5294 &11799\\
         \midrule
        \textbf{Avg Record Per Day, \(\bar{r}\) } & \multicolumn{6}{c}{9456}\\
         \bottomrule
    \end{tabular}
    \vspace{-2mm}
\end{table}

\begin{figure}[!h]
    \vspace{-3mm}
    \centering
    \includegraphics[width=1.0\linewidth]{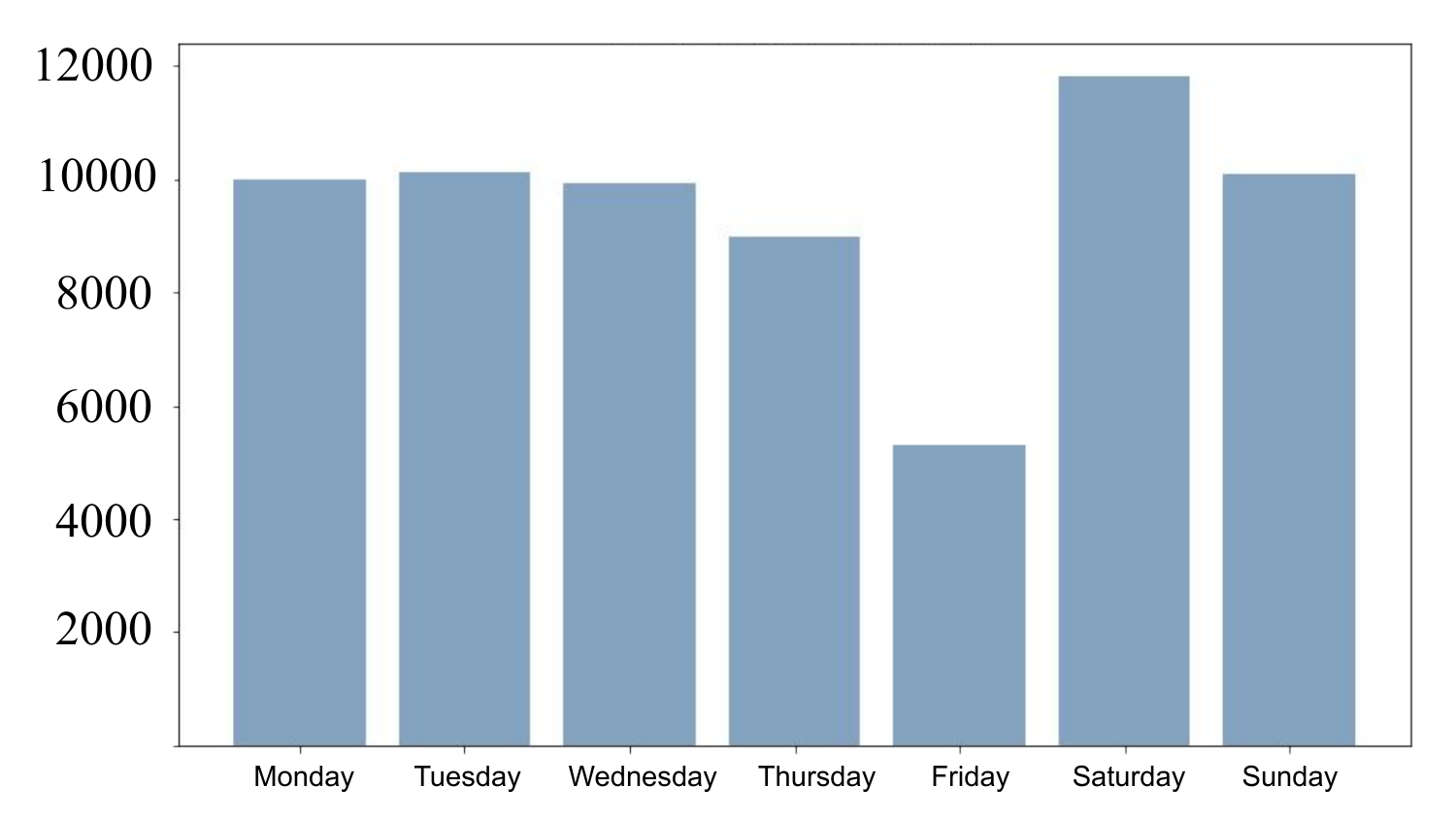}
    \caption{Day-Wise Weekly Entry Count}
    \label{fig:Day-wise Weekly Entry Count}
\end{figure}

Table~\ref{tab:daily-avg-record} provides average daily
entries (\(r\)) for a single hospital or diagnostic center. Saturday
and Sunday generally show higher traffic, while Friday tends
to be lowest due to reduced staffing , as shown in Fig.~\ref{fig:Day-wise Weekly Entry Count}. From Table~\ref{tab:daily-avg-record}, the average per-hospital/center daily record is \(\bar{r} = 9456\). Scaling to the national level requires accounting for 552 government hospitals (secondary/tertiary), 8000 private diagnostic centers, and different “weights” (\(w\)) based on each hospital’s bed capacity.


Table~\ref{tab:seat-hospital-weight} shows how hospital bed-capacity weights translate into daily record estimates. Thus, roughly \(19.04\) million records flow into the NCDW each day.

\begin{table}[!h]
\centering
\caption{Load estimation of dfferent Hospital.}
\label{tab:seat-hospital-weight}
\begin{tabular}{c|c|c|c}
\toprule
\textbf{Seat No} (\(s\)) & \textbf{Hospital Count} (\(n\)) & \textbf{Weight} (\(w\)) & \(\textbf{w} \times n \times \bar{r}\) \\
\hline
10 & 17 & 0.0030 & 486 \\
20 & 32 & 0.0060 & 1828 \\
31 & 266 & 0.0093 & 23551 \\
50 & 158 & 0.0151 & 22562 \\
100 & 31 & 0.0302 & 8854 \\
150 & 1 & 0.0453 & 429 \\
200 & 1 & 0.0604 & 572 \\
250 & 26 & 0.0755 & 18564 \\
500 & 2 & 0.1510 & 2856 \\
500 & 11 & 0.1510 & 15708 \\
1500 & 7 & 0.4530 & 29988 \\
\hline
\multicolumn{3}{c|}{\textbf{Total for 552 Govt. Hospitals}} & 125398 \\
\hline
\multicolumn{3}{c|}{\textbf{Diagnostic Centers (8000)}} & 18912000 \\
\hline
\multicolumn{3}{c|}{\textbf{Overall Daily Total (8552)}} & 19037398 \\
\bottomrule
\end{tabular}
\vspace{-2mm}
\end{table}

Now assuming each record requires \(0.1\,\mathrm{KB}\), the estimated daily storage usage is given 
\(\mathrm{Daily\ Size} \approx R \times 0.1\,\mathrm{KB}\) according to Eq. ~\ref{eqn2}. 
Table~\ref{table:size-estimate} presents projected storage volumes over different periods. 
The daily storage requirement is approximately \(19\,\mathrm{GB}\), increasing to nearly \(7\,\mathrm{TB}\) annually and \(34\,\mathrm{TB}\) over five years.

\begin{table}[!h]
\centering
\caption{Estimated size of NCDW for different time spans.}
\label{table:size-estimate}
\begin{tabular}{c|c|c}
\toprule
\textbf{Total Records/Day, \(R\)} & \textbf{Time Span, \(d\)} & \textbf{Estimated Size} \\
\hline
\multirow{3}{*}{19,037,398} 
  & 1 day & 19.04 GB  \\ \cline{2-3}
  & 1 year (365 days) & 6.79 TB  \\ \cline{2-3}
  & 5 years (5$\times$365 days) & 33.95 TB  \\ 
\bottomrule
\end{tabular}
\vspace{-2mm}
\end{table}


    \begin{figure*}[!t]
        \centering
        \includegraphics[width=1.0\textwidth,trim={0 150 0 60}, clip]{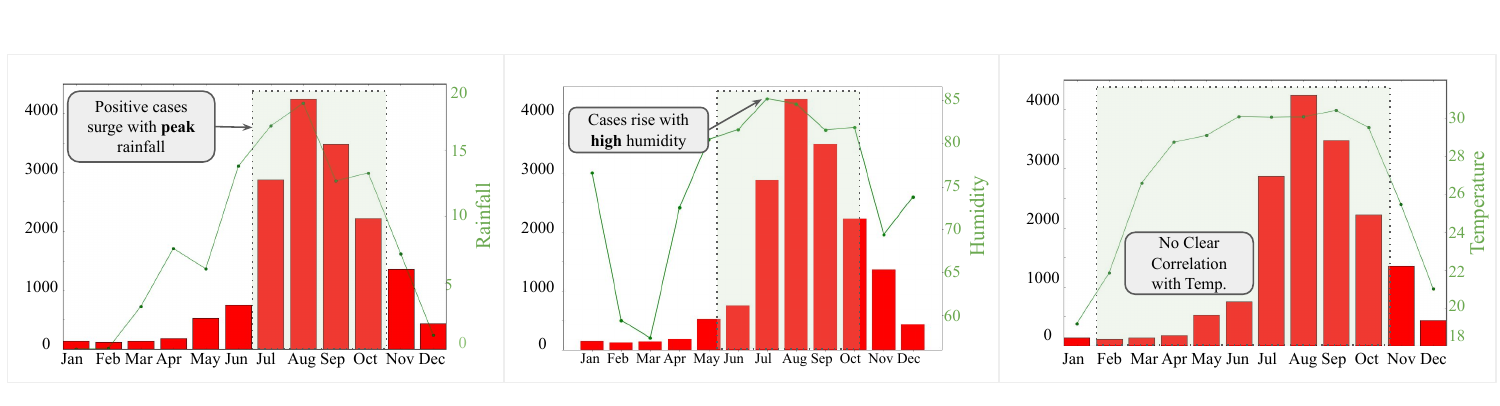}
        \vspace{-15pt}
        \begin{center}
           (a) Rainfall vs Positive Cases \hspace{1.5cm} 
           (b) Humidity vs Positive Cases \hspace{1.5cm} 
           (c) Temperature vs Positive Cases
        \end{center}
        \vspace{-4pt}
        \caption{Impact of environmental factors on dengue cases. Rainfall (a) and humidity (b) exhibit a significant correlation with dengue cases, whereas temperature (c) shows no clear correlation.}
        \label{fig:rainfallvshum}
        \vspace{-5mm}
    \end{figure*}


\subsubsection{Runtime and Performance Comparison (PostgreSQL vs. HBase) }
To evaluate the performance of PostgreSQL and HBase for the NCDW, cube operations were conducted on datasets of varying sizes (100K, 200K, 500K, and 1M), and the execution times for cube sizes 3 and 4 were measured for both systems, as summarized in Table~\ref{tab:cube34}.

\begin{table}[!h]
\centering
\caption{Execution times (in minutes) for cube sizes 3 and 4.}
\begin{tabular}{c|c|c|c|c}
\toprule
\multirow{2}{*}{\textbf{Aggregation Size}} & 
\multicolumn{2}{c|}{\textbf{Cube Size 3}} & 
\multicolumn{2}{c}{\textbf{Cube Size 4}} \\ 
 & \textbf{PostgreSQL} & \textbf{HBase} & \textbf{PostgreSQL} & \textbf{HBase} \\ 
 \hline
100,000   & 5    & 3    & 92   & 40   \\ \hline
200,000   & 49   & 15   & 150  & 70   \\ \hline
500,000   & 150  & 60   & 350  & 180  \\ \hline
1,000,000 & 400  & 180  & 780  & 410  \\ 
\bottomrule
\end{tabular}
\label{tab:cube34}
\vspace{-2mm}
\end{table}

\begin{figure}[h]
    \centering
    \includegraphics[width=0.5\textwidth,trim={10 60 0 70}, clip]{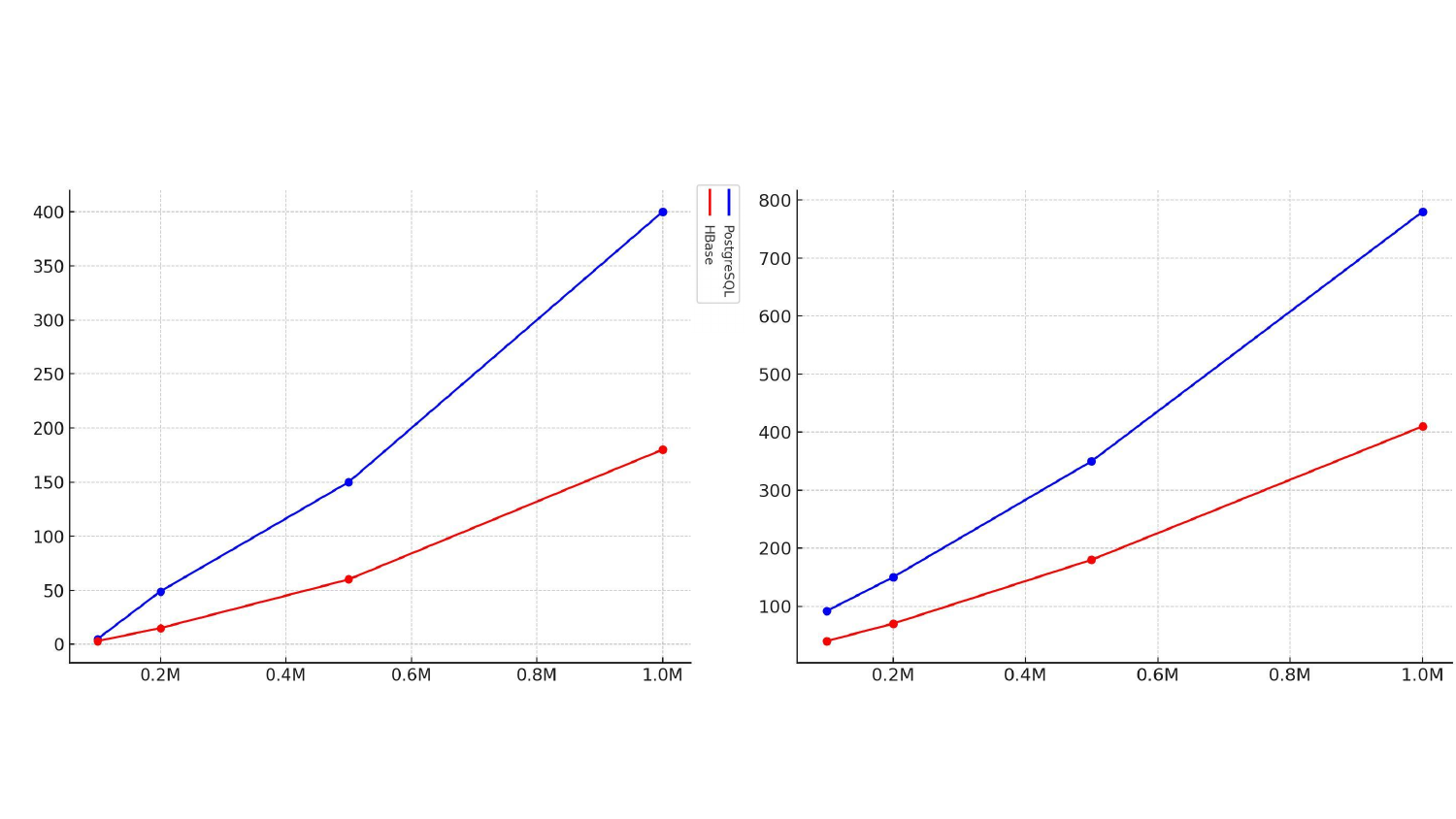}
    \begin{center}
       (a) Cube Size 3 \hspace{1.9cm} (b) Cube Size 4
    \end{center}
    \vspace{-4pt}
    \caption{Execution time comparison for cube size 3 (a) and cube size 4 (b) between PostgreSQL and HBase.}
    \label{fig:posthbs}
\end{figure}

Fig.~\ref{fig:posthbs} illustrates that HBase outperforms PostgreSQL, with execution times improved by 40–69\% across different cube and dataset sizes.

\begin{figure}
\vspace{-4mm}
    \centering
    \includegraphics[width=0.9\linewidth]{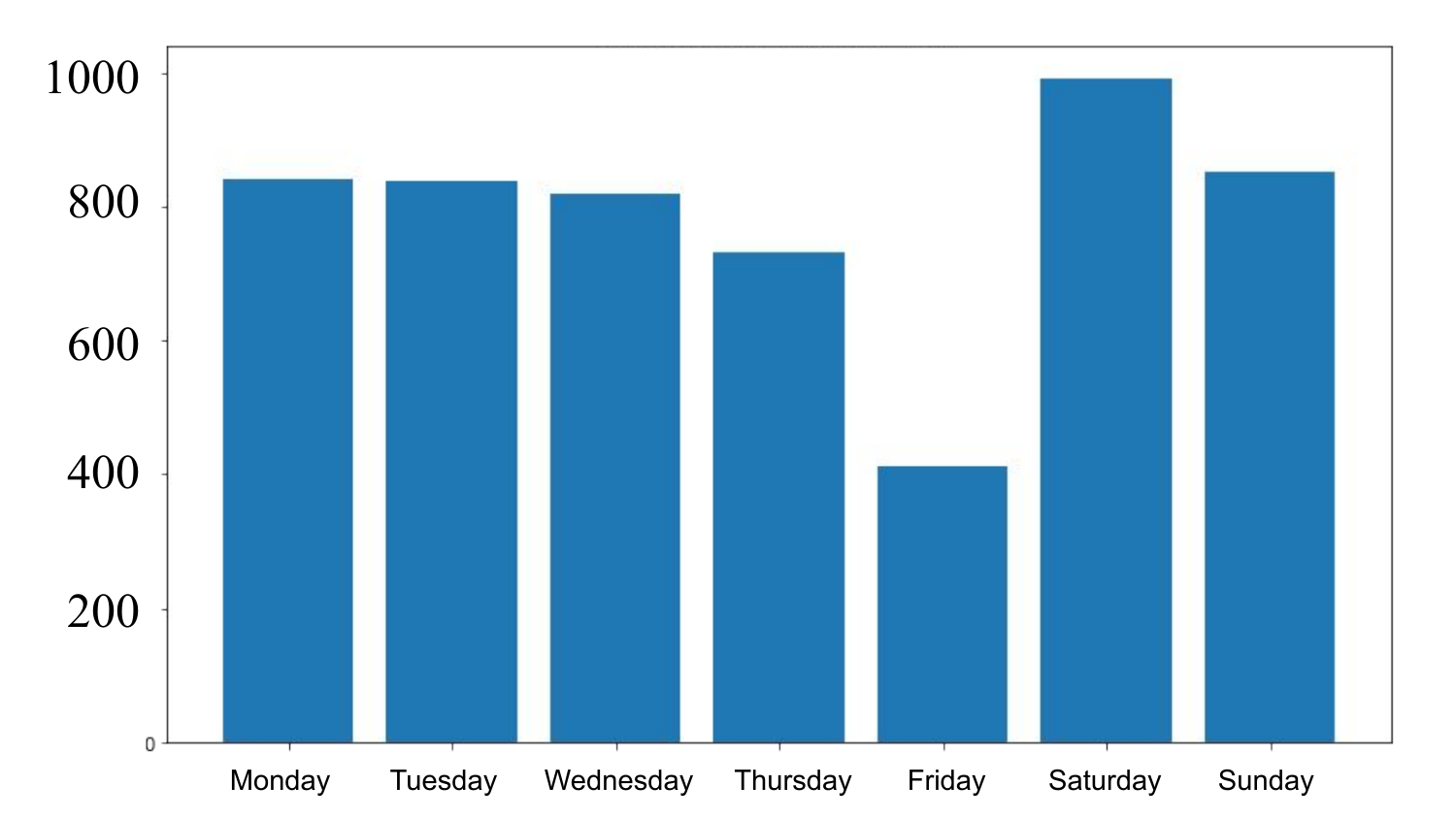}
    \caption{Day-wise Weekly Test Count(Avg)}
    \label{fig:day_wise_weekly_test}
    \vspace{-6.5mm}
\end{figure}

\subsection{ Data  Mart Specific Insights}
The second highlights the dengue data mart case study, demonstrating how the NCDW derives epidemiological insights from integrated health and environmental data.

\subsubsection{Environmental Correlation with Dengue Cases}
Environmental factors such as rainfall, humidity, and temperature play a critical role in dengue incidence. Rainfall fosters mosquito breeding by creating stagnant water, with dengue cases peaking in August, coinciding with the highest rainfall levels, as shown in Fig.~\ref{fig:rainfallvshum}a. Humidity also exhibited a strong correlation with dengue cases, reaching its peak in August, as illustrated in Fig.~\ref{fig:rainfallvshum}b, suggesting its potential as an early outbreak indicator. In contrast, temperature fluctuations had no significant impact on dengue incidence, as demonstrated in Fig.~\ref{fig:rainfallvshum}c. These insights highlight the importance of environmental monitoring for effective dengue prevention and control.

\begin{figure}[!h]
    \centering
    \includegraphics[width=0.9\linewidth,trim={0 50 0 50},clip]{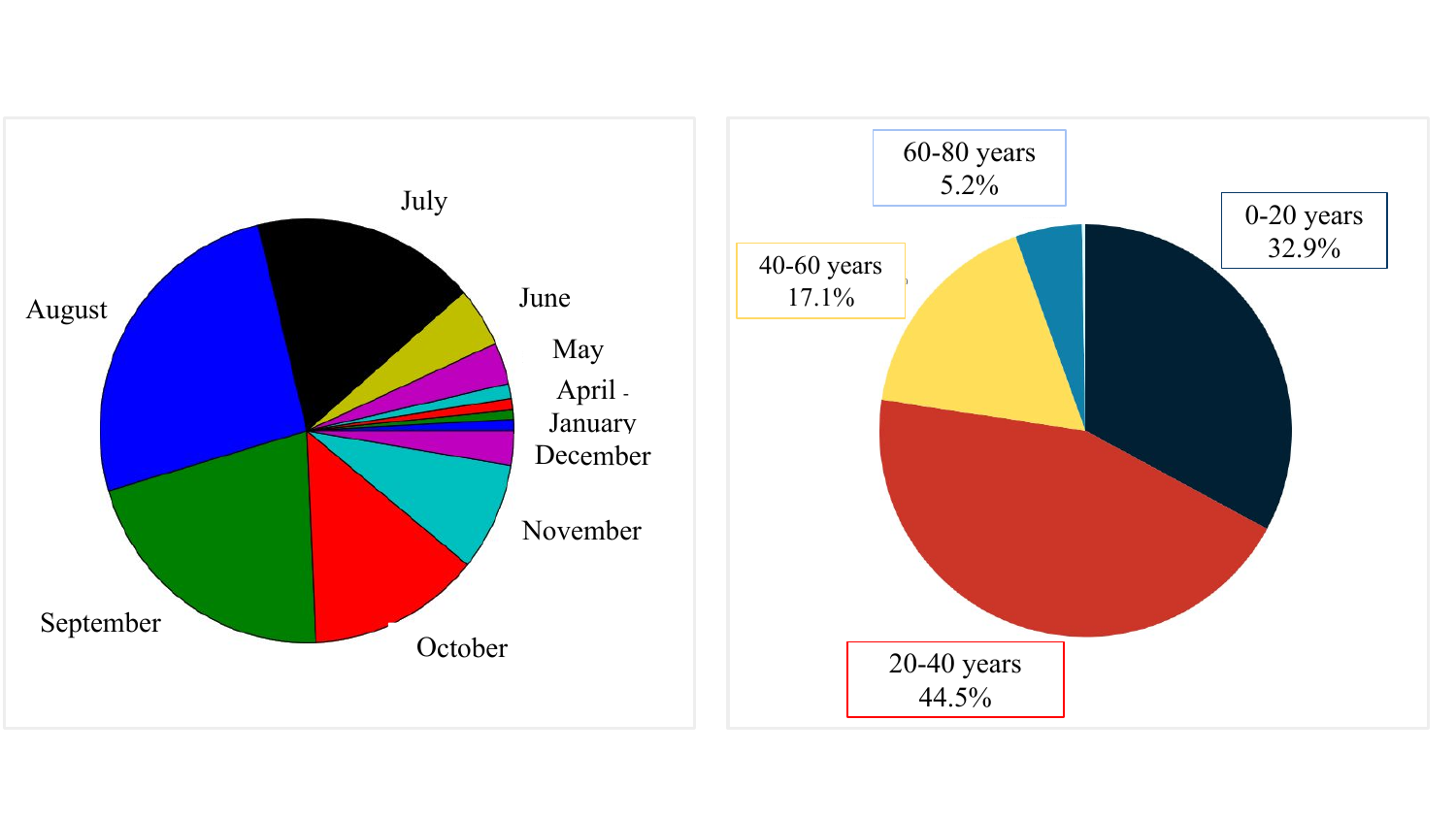}
    \begin{center}
     (a) Monthly distribution \hspace{1cm}  (b) Age-wise distribution
    \end{center}
    \caption{Dengue-positive cases by month and age group. (a) Peak incidence from July to October. (b) Higher vulnerability among individuals aged 0–40 years.}

    \label{fig:demographic}
\end{figure}

\subsubsection{Temporal and Demographic Trends of Dengue Incidence}
Dengue cases followed a seasonal pattern, peaking between July and October, with the highest incidence in August, aligning with the monsoon season and increased mosquito breeding, as shown in Fig.~\ref{fig:demographic}a. Weekly trends indicate peak data entry on Saturdays and Sundays, with a noticeable drop on Fridays, likely due to staffing variations, as illustrated in Fig.~\ref{fig:day_wise_weekly_test}. Additionally, 77.4\% of cases were reported in the 0–40 age group, emphasizing the need for targeted prevention efforts, as depicted in Fig.~\ref{fig:demographic}b.
\begin{figure}[]
    \vspace{-3.5mm}
    \centering
    \includegraphics[width=0.95\linewidth]{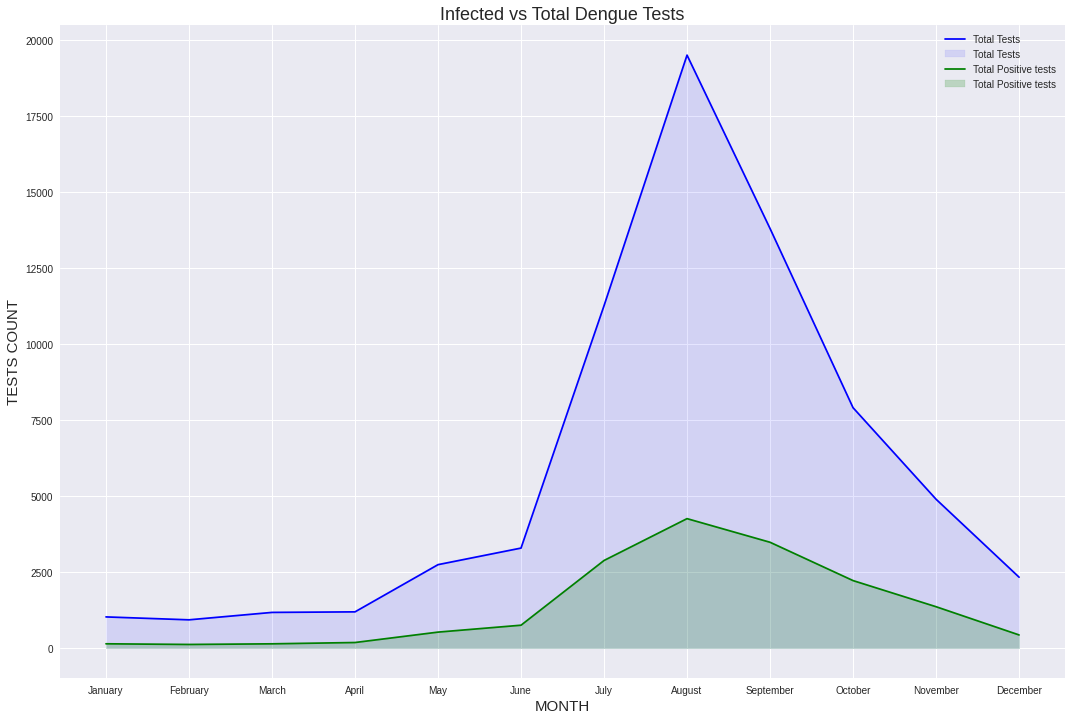}
    \caption{Outbreak Detection}
    \label{fig:outbreak-detection}
    \vspace{-6mm}
\end{figure}
\subsubsection{Outbreak Detection}
Tracking monthly case counts enabled early detection of a surge in infections beginning in June, reaching its highest point in August, and declining thereafter, as shown in Fig.~\ref{fig:outbreak-detection}. These findings confirm a July–October outbreak period and highlight the system’s capability to identify such surges in near-real time, supporting proactive healthcare measures.

%% file: tex_generalized/limitations.tex
\section{Limitations and Future Work}
Although the proposed NCDW framework is designed for scalability, its effectiveness depends on the consistency and availability of data sources. The limited inclusion of behavioral or personal patient traits poses a challenge in generating comprehensive patient insights. Future research should explore methods to incorporate patient lifestyle data to enhance predictive analytics and disease surveillance.

Additionally, while our framework enables efficient disease-specific analytics, it currently does not support complex data types such as test reports and medical imaging (e.g., X-rays, MRIs). Expanding the system’s storage capabilities to accommodate such data will be essential for holistic patient record management.

Another key area for improvement is the database architecture. Our SQL vs. NoSQL comparison revealed significant performance differences, suggesting that a deeper investigation into NoSQL-based solutions could optimize the handling of large-scale EHR data. Further research should evaluate hybrid approaches that balance structured (clinical records) and semi-structured (environmental, genomic) data storage. 

By addressing these challenges, the NCDW framework can evolve into a more robust, flexible, and predictive system, supporting evidence-based public health strategies and efficient resource allocation across diverse healthcare environments.

%% file: tex_generalized/conclusion.tex
\section{Conclusion}
Our study presents a framework for a National Clinical Data Warehouse (NCDW) that is designed to address the scalability, interoperability, and privacy challenges inherent in resource-constrained healthcare systems. The proposed wrapped-based architecture ensures secure and anonymized data integration, while our proposed name-matching algorithm effectively resolves patient identity inconsistencies in the absence of unique identifiers.

To evaluate the feasibility and impact of the proposed framework, we conducted a case study using healthcare data from Bangladesh. Our prototype deployed in our institution's server estimated processing 19 million daily records across 8552 hospitals. Additionally, the dengue-specific data mart revealed a clear correlation between dengue outbreaks and environmental factors such as rainfall and humidity, showcasing the potential of NCDW-driven analytics in epidemiological surveillance and outbreak prediction. 

While our case study is focused on Bangladesh, the framework is designed to be adaptable to other developing nations facing similar challenges. By incorporating secure data acquisition and smart storage solutions, our approach provides a generalizable model that can be tailored to different national healthcare infrastructures. Our proposed star schema can also be adapted to incorporate more data types (i.e., patient demographics,
MRI/X-ray images). Future work should focus on expanding this framework to support additional diseases, integrating predictive models for outbreak forecasting, and refining storage solutions to accommodate imaging and genomic data. By implementing this framework, developing nations can enhance their public health decision-support systems, optimize resource allocation, and improve patient care on a national scale.

